# Effect of Crosstalk on Permutation in Optical Multistage Interconnection Networks

Er.Sandeep Kaur, Er.Anantdeep and Er.Deepak Aggarwal

**Abstract** — Optical MINs hold great promise and have advantages over their electronic networks.they also hold their own challenges. More research has been done o n Electronic Multistage Interconnection Networks, (EMINs) but these days optical communication is a good networking choice to meet the increasing demands of high-performance computing communication applications for high bandwidth applications. The electronic Multistage Interconnection Networks (EMINs) and the Optical Multistage Interconnection Networks (OMINs) have many similarities, but there are some fundamental differences between them such as the optical-loss during switching and the crosstalk problem in the optical switches. To reduce the negative effect of crosstalk, various approaches which apply the concept of dilation in either the space or time domain have been proposed. With the space domain approach, extra SEs are used to ensure that at most one input and one output of every SE will be used at any given time. For an Optical network without crosstalk, it is needed to divide the messages into several groups, and then deliver the messages using one time slot (pass) for each group, which is called the time division multiplexing. This Paper discusses the permutation passability behavior of optical MINs. The bandwidth of optical MINs with or without crosstalk has also been explained. The results thus obtained shows that the performance of the networks improves by allowing crosstalk to some extent.

—————————— ◆ ——————————

## 1 INTRODUCTION

Optical communication is necessary for achieving reliable, quick and flexible communication. Advances in optical technologies have made optical communication a reliable networking choice to meet the demands for high bandwidth and low communication latency of high-performance computing/communication applications. So optical networks gives high performance as well as low latency .Although optical MINs hold great promise and have advantages over their electronic networks, they also hold their own challenges. Advances in electro-optic technologies have made optical communication a good networking choice for the increasing demands of high channel bandwidth and low communication latency of high-performance computing/communication applications. Fiber optic communications offer a combination of high bandwidth, low error probability, and gigabit transmission capacity. Multistage Interconnection Networks (MINs) are very popular in switching and communication applications and have been used in telecommunication and parallel computing systems. But these days with growing demand for bandwidth, optical technology is used to implement interconnection networks and switches. In electronic MINs electricity is used, where as in Optical MINs (OMIN) light is used to transmit the messages [21]. The electronic MINs and the optical MINs have many similarities, but there are some fundamental differences between them such as the optical-loss during switching and the crosstalk problem in the optical switches. Optical interconnections have the potential of becoming an appealing alternative to electrical interconnections. For long and medium range distances (e.g., local area networks and telecommunication), optical technology (fibers) is the technology of choice, offering better performance and lower costs than electrical wires [21]. There is a trend for optics to replace electronics for shorter distances and larger connectivity applications.Optical interconnections are insensitive to radio wave interference effects, are free from transmission line capacitive loading, do not have geometrical planar constraints, and can be reconfigurable (circuit-switched)[6,7].Crosstalk in optical networks is one of the major shortcomings in optical switching networks, and avoiding crosstalk is an important for making optical communication properly. To avoid a crosstalk, many approaches have been used such as time domain and space domain approaches [22]. Because the messages should be partitioned into several groups to send to the network, some methods are used to find conflicts between the messages.

## 2 OPTICAL MULTISTAGE INTERCONNECTION NETWORKS

An optical MIN can be implemented with either free-space optics or guided wave technology. It uses the Time Division Multiplexing. To exploit the huge optical bandwidth of fiber, the Wavelength Division Multiplexing (WDM) technique can also be used. With WDM, the optical spectrum is divided into many different logical channels, and each channel corresponds to a unique wavelength. Optical switching, involves the switching of optical signals, rather than electronic signals as in conventional electronic systems. Two types of guided wave optical switching systems can be used. The first is a hybrid approach in which optical signals are switched, but the switches are electronically controlled .With this approach, the use of electronic control signals means that the routing will be carried out electronically. As such, the speed of the electronic switch control signals can be much less than the bit rate of the optical signals being switched.So, with this approach there is a big speed mismatch occurs due to the high speed of optical signals. The second approach is all- optical switching. This has removed the problem that occurred with the hybrid approach. But, such systems will not become practical in the future and hence only hybrid optical MINs are considered. In hybrid optical MINs, the electronically controlled optical switches, such as lithium neonate directional couplers, can have switching speeds from hundreds of picoseconds to tens of nanoseconds.

## 3 PROBLEMS

Path dependent loss means that optical signals become weak after passing through an optical path. In a large MIN, a big part of the path-dependent loss is directly proportional to the number of couplers that the optical path passes through. Hence, it depends on the architecture used and its network size. Hence, if the optical signal has to pass through more no of stages or switches, the path dependent loss will be more.

Optical crosstalk occurs when two signal channels interact with each other. There are two ways in which optical paths can interact in a switching network. The channels carrying the signals could cross each other.The second way is when two paths sharing a switch could experience some undesired coupling from one path to another within a switch.

## 4 SOLVE CROSSTALK

One way to solve the crosstalk problem is a space domain approach, where a MIN is duplicated and combined to avoid crosstalk. The number of switches required for the same connectivity in a network with space domain approach is slightly larger than twice that for the regular network. This approach uses more than double the original network hardware to achieve the same. Thus for the same permutation the hardware or we can say the no of switches will be double. Thus cost will be more with the networks using space domain approach. In the entire four cases only one input and only one output is active at a given time so that no cross talk occurs. With the space domain approach, extra switching elements (SEs) (and links) are used to ensure that at most one input and one output of every SE will be used at any given time.

## 5 PERMUTATION PASSABILITY

Permutation passability means how many input requests occurring simultaneously at the input are able to pass through a given network and how many of them will successfully mature i.e. will reach their destination [21]. The request always pass from the most suitable path available (generally, the minimum length path), if such path is busy or faulty then the request is pass through an alternate path. If no alternate path is available then the request has to be simply dropped or said to be having clash. So some of the requests will pass through the most favorable path, others have to be routed through an available alternative path. If no alternative paths are available then some requests cannot be served at all. Crosstalk in optical networks is one of the major shortcomings in optical switching networks, and avoiding crosstalk is an important for making optical communication properly. To avoid a crosstalk, many approaches have been used such as time domain and space domain approaches. Because the messages should be partitioned into several groups to send to the network, some methods are used to find conflicts between the messages. If we allow the limited crosstalk in the optical networks the permutation passibility of the optical networks will be increased [21].

## 6 CALCULATION AND DISCUSSION

### 6.1 No of passes of Banyan Network

To avoid Crosstalk only one input is allowed to pass through one switch For Example In Banyan network if all the 8 inputs get active only 4 inputs would be allowed to pass through the network towards its output to avoid the problem of cross talk.

    Inputs     0 1 2 3 4 5 6 7
    Output     7 0 5 2 3 6 1 4

It is decomposed into two permutations in the first Pass.

    Inputs    0 1 2 3      and    Inputs  4 5 6 7
    Output    7 0 5 2             Output  3 6 1 4

These two passes would solve the crosstalk problem at the first and last stage, the Intermediate stage would still be having the problem of crosstalk. Thus the permutations again has to break to avoid the crosstalk problem. This can be done with the decomposition again. It can be possible with the bipartite graph. Crosstalk is the dangerous problem for the banyan Network. This should be removed.

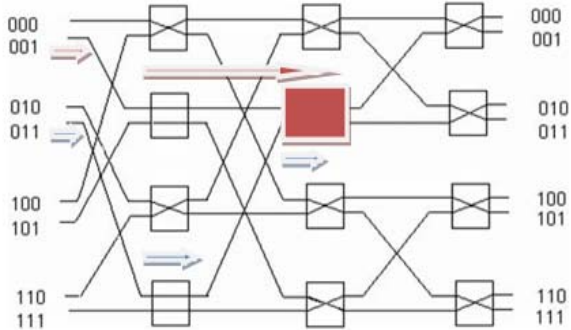

Figure 1(a): Crosstalk at the Middle stage in Pass 1

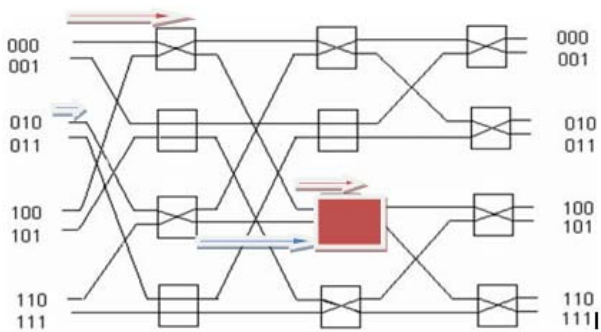

Figure 1(b): Crosstalk at the Middle stage in Pass1

To avoid this problem (crosstalk at the intermediate stage), we divide the permutations into semi permutations. Semi permutation of pass 1 in the given example is

Input     0 1     and     2 3
Output    7 0              5 2

Second pass permutation will be
Input     4 5     and     6 7
Output    3 6              1 4

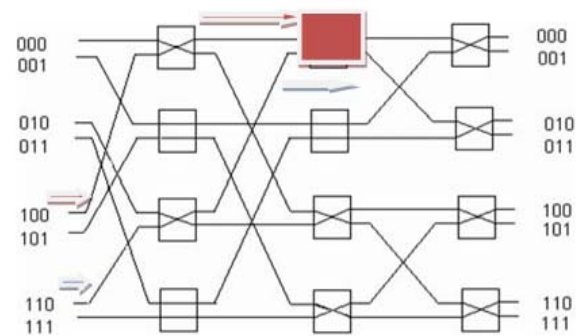

Figure 1(c): Crosstalk at the Middle stage in Pass2

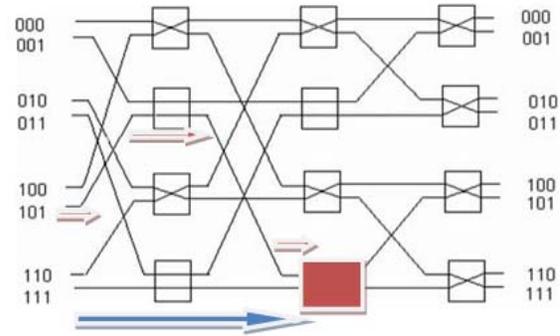

Figure 1(d): Crosstalk at the Middle stage in Pass2

### 6.2 Bandwidth Comparison

The bandwidth of stage banyan network ($N = 2^n$ is the network size) is given by:-
BW = P (n) ×Size of Network [45]

The value of $P(n)$ can be obtained from the probabilistic Equations In this the effect of Crosstalk is studied on the Permutation of Optical MINS. Optical multistage interconnection networks (OMINS), which interconnect their inputs and outputs via several stages of switching elements using optical guided waves or free space, is studied. Although optical MINS hold great promises and have demonstrated advantages over their electronic counterparts, they also introduce new challenges and problems of avoiding cross-talks in the switching elements [45].

In this the effect of limited Crosstalk is studied on the Permutation of Optical MINS. Optical multistage interconnection networks (OMINS), which interconnect their inputs and outputs via several stages of switching elements using optical guided waves or free space, is studied. Although optical MINS hold great promises and have demonstrated advantages over their electronic counterparts, they also introduce new challenges and problems of avoiding cross-talks in the switching elements. Interact with each other. There are two ways in which optical paths can interact in a switching network. The channels carrying the signals could cross each other and When the two paths sharing a switch could experience some undesired coupling from one path to another within a switch. Crosstalk problem is more dangerous than the path-dependent loss problem with current optical technology. Thus, switch crosstalk is the most significant factor that reduces the signal-to-noise ratio and limits the size of a network. Luckily, first-order crosstalk can be eliminated by ensuring that a switch is not used by two input signals simultaneously. Once the major source of crosstalk disappears, crosstalk in an optical MIN will have a very small effect on the signal-to-noise ratio and thus a large optical MIN can be built and effectively used in parallel computing systems. Initially it is checked that bandwidth of Optical Networks with crosstalk is greater than without crosstalk networks for Banyan and Baseline network. The bandwidth is calculated in terms of probability of whether the switch is active or not using the probability equations. The bandwidth without crosstalk and with crosstalk is calculated.

**Table 1: Banyan Network Bandwidth**

| Size | Bandwidth |
| --- | --- |
| 4 | 1.9 |
| 8 | 3.81 |
| 16 | 6.2 |
| 32 | 12.1 |
| 64 | 22 |

**Table 2: Bandwidth with allowing crosstalk to some extent**

| Size | Bandwidth |
| --- | --- |
| 4 | 1.9 |
| 8 | 6 |
| 16 | 10 |
| 32 | 16 |
| 64 | 54 |

## 7 CONCLUSION

In this, we analyzed various advantages of Optical Networks over the Electronic Networks. So, we conclude that for today's applications such as in WAN'S Optical networks are the promising choice to meet the high demand of Speed and Bandwidth. We concluded that optical loss can be removed by using the space and time-space approach.Bandwidths of OMINs have been analyzed with and without crosstalk. It has been observed that OMINs provide higher bandwidth with some crosstalk than without crosstalk.

## 8 FUTURE WORK

Bandwidth can be calculated under Traffic and Bursty conditions.Performance can be calculated for OMINs that do not have self-routing capabilities.Permutations Capability of more networks with or without faults can be checked.

### ACKNOWLEDGMENT

We are greatful to Er.Rinkle Aggarwal, Senior lecturer, Thapar University, Patiala for her constant encouragement that was of great importance in the completion of the paper.

### REFERENCES

[1] Adams George B, Agrawal Dharma P, Siegel Howard Jay, "A Survey and Comparison of Fault-Tolerant Interconnection Networks", IEEE Transactions on Computers, pp.14-27, June 1987.
[2] A.Subramanyam, E.V.Prasad, C.Nadamuni Reddy, "Performance of Processor Memory interconnection Networks", proceedings of 2nd Natnl, "NCMCM – 2003", pp 389 - 394, June 2003.
[3]A. Himeno, and M. Kobayashi, "4 _ 4 optical-gate matrix switch", J. Light wave Technology, vol. 3, pp. 230-235, April 1985.
[4]Bhuyan, L.N. Agrawal, D.P, "Design and performance of generalized interconnection networks", IEEE Transactions on Computers, vol. 32, pp. 1081-1090, 1983.
[5]Benjamin A. Small, Keren Bergman, "Optimization of Multiple-Stage Optical Interconnection Networks", IEEE Photonics Technology Letters, vol. 18, no. 1, January 2006
[6]Brenner M, Tutsch D, Hommel G, "Measuring transient performance of a multistage interconnection network using ethernet networking equipment", Proceedings of the International Conference on Communications in Computing 2002 (CIC'02), pp. 211–216, Las Vegas, USA, 2002
[7]Bhuyan Laxmi N, Yang Qing, Aggarwal P Dharma, "Performance of Multiprocessor Interconnection Networks", Proceeding of IEEE, pp. 25-37, February 1989.
[8]Blaket James T, Trivedi Kishor S, "Reliabilities of Two Fault-Tolerant Interconnection Networks", Proceeding of IEEE, pp. 300-305, 1988.
[9]Chuan-Lin Wu, Tse-Yun Feng, "The Reverse-Exchange Interconnection Network", IEEE Transactions on Computers, vol. c-29, no. 9, pp. 801-811, September 1980.
[10]Chi Hsin-Chou, Wu Wen-Jen, "Routing Tree Construction for Interconnection Networks with Irregular Topologies", Proceeding of the Eleventh Euromicro Conference on Parallel, Distributed and Network-Based Processing (Euro-PDP), 2003.
[11]C. Qiao, "A two-level process for diagnosing crosstalk in photonic dilated Benes network", Journal of Parallel and Distributed Computing, vol. 41, no. 1, pp. 53-66, 1997.
[12]C.Wu, T. Feng, "On a class of multistage interconnection networks", IEEE Transactions on. Computers, vol. c-29, pp. 694–702, August 1980.
[13]D. K. Hunter, I. Andonovic, "Guided wave optical switch architectures", International Journal of Optoelectronics, vol. 9, no. 6, pp. 477-487, 1994.
[14]D.Nassimi, S.Sahni, "A self routing benes network and parallel permutation algorithms", IEEE Transactions on computers, vol. c-30 (5), pp. 332-340, 1981.
[15]D. Cantor, "On nonblocking switching networks", vol. 1, 1971.
[16]D. K. Pradhan, K. L. Kodandapani, "A uniform representation of single and multistage interconnection networks used in SIMD machines," IEEE Transactions on Computers. vol. 29, pp. 777–791, Sept. 1980.
[17]E.V.Prasad, A.K.Sarjee "Performance Evaluation of Multiprocessor System Modelled as t-out-of-Systems", International of Journal of High Speed Computing, vol. 5, no.1, pp 91-87, 1993.
[18]Goke, L.R, Lipovski, "Banyan networks for partitioning multiprocessing systems", International: Proceedings of the First International Symposium on Computer Architecture, pp. 21-28, 1973.
[19]H. Hinton, "A non-blocking optical interconnection network using directional couplers", Proc. IEEE, Global Telecommunications Conference, pp. 885-889, November 1984.
[20] Hao Tian, Yi Pan, Ajay K Katangur, Jiling Zhong, "A novel modularized Optical Multistage Interconnection architecture with multicast capability", Department of Computer Science Georgia State University, Atlanta, GA 30303.
[21]J ianchao Wang, YiPan," Permutation Capability of Optical Multistage Interconnection Networks", International and Symposium on Parallel and Distributed Processing, Parallel processing Symposium, April 1998.

[22] K.M. Sivalingam, S. Subramaniam, "Optical WDM Networks, Principles and Practice", Kluwer Academic Publishers, 2000.

[23] K.V Arya, R.K. Ghosh "Designing a New Class of Fault Tolerant Multistage Interconnection Networks", Journal of Interconnection Networks, vol. 6, No 4, pp. 361-382, 2005.

[24] K. Padmanabhan, A.N. Netravali, " Dilated networks for photonic switching", IEEE Transactions on.Communications, vol. 35,no. 12, pp. 1357-1365, December 1987.

[25] Mittal R, Cherian D, Mohan P.J, " Routing and Performance of the double tree (DOT) network", Proceeding of International Conference on Computer Digital Technology, vol. 142, no. 2, pp. 93-97, March 2005.

[26] M.Xu, S.Shen, " A new fault-tolerant generalized cube with an extra stage", Proc.Intl. Comp.Sc.Conf.(ICSC'92), pp. 99-105, December 1992.

[27] "Optical Multistage Interconnection Networks: New Challenges and Approaches", IEEE Communications Magazine, February 1999.

[28] Othman M, Abedi F, "Fast Method to find conflicts in Optical Multistage Interconnection Networks", International Journal of The Computer, The Internet and Management ,vol. 16, pp. 18-25, April 2008.

[29] Prasad A.R, Reddy E.V, "Permutation Capability and Connectivity of Enhanced Multistage Interconnection", **ADCOM 2006, International Conference on** Advanced Computing and Communications, pp. 8-11, December 2006

[30] Qiao, C.Zhou, "Scheduling switching element disjoint connections in a stage controlled photonic banyans", IEEE Transactions on.Communications, vol.47, pp. 139-148, 1999.

[31] Qiao, C, Melhem, R, Chiarulli, D, Levitan S, "A time domain approach for avoiding crosstalk in optical blocking multistage interconnection networks", J. Light. Technology, pp. 1854-1862, 1994.

[32] R.A. Thompson, "The dilated slipped banyan switching network architecture for use in an all-optical local-area network", J. Lightwave Technology, vol. 9, no. 12, pp. 1780-1787, December 1991.

[33] Regis Bates, J. "Optical Switching and Networking", Handbook.McGraw-Hill, New York, 2001.

[34] Ren Kaixin, G.U Naijie, "Permutation Capability of Optical Cantor Network", Eighth International Conference on Parallel and Distributed Computing, Applications and Technologies, 2007.

[35] Shen, X., Yang, F., Pan, Y.: "Equivalent permutation capabilities between time division optical omega networks and non-optical extra-stage omega networks." IEEE/ACM Transactions on Networks **9**(4), pp. 518–524, 2001.

[36] Sengupta J, Bansal P.K, "Performance of Regular and Irregular Dynamic MINs", Proceeding of International Conference IEEE Tencon, pp. 427-430,1999.

[37] Sengupta J, Bansal P.K, Gupta Ajay, " Permutation and Reliability measures of Regular and Irregular MINs", International Conference IEEE, pp. I-531-I-536, 2000.

[38] Tse-yun Feng, Seung-Woo Seo, "New Routing Algorithm for a Class of Rearrangeable Networks", IEEE Transactions on computer, vol. 43, No. 11, November 1999.

[39] Vaez M.M, Lea C.T, "Strictly nonblocking directional-couplerbased switching networks under crosstalk constraint", IEEE Transactions on Communications, vol. 48, pp. 316-323, 2000.

[40] Varma A, Raghavendra C.S, "Interconnection Networks for Multiprocessors and Multicomputers", Theory and Practice, IEEE Computer Society Press, Los Alamitos ,1994.

[41] Wu Xingfu, Sun Xian-He,"Performance Modeling for Interconnection Networks", Proceeding of International Conference IEEE, pp. 380-385, 2000

[42] Xiaohong Jiang, Hong Shen,"A New Scheme to Realize Crosstalk-free Permutations in Optical MINs with Vertical Stacking", Proceedings of the International Symposium on Parallel Architectures,pp. 341,2002

[43] Yuanyuan Yang, Jianchao Wang "Permutation Capability of Optical Multistage Interconnection Networks", Journal of Parallel and Distributed Computing, vol. 60, pp. 72-91, 2000.

[44] Y.Yeh, T.Feng, "On a class of rearrangeable networks", IEEE Transactions on computers, vol. 141 (11), pp 1361-1379, 1992.

[45] Yi Pan, Ajay K Katangur, Somasheker, "Analyzing the performance of optical multistage interconnection networks with limited crosstalk",Kluwer Academic Publishers, vol. 10, 2007.

[46] Yuanyuan Yang,Jianchao Wang,"Optimal All-to-All Personalized Exchange in a Class of Optical Multistage Networks", IEEE Transactions on Parallel and Distributed Systems,vol 12 ,pp. 567 - 582, june 2001.

[47] Zheng S.Q, Lu E, "High-speed crosstalk-free routing for optical multistage interconnection networks", Computer Communications and Networks, ICCCN 2003, Proceedings on the 12th International Conference on Networks, pp. 249- 254, October 2003.

**Er. Sandeep Kaur**, presently working as Lecturer in the Department CSE/IT of BBSB Engineering College, Fatehgarh Sahib, Punjab (INDIA). She is master of engineering in computer science & engineering from Thapar University Patiala. Her major research interests include parallel computing and performance of optical multistage interconnection networks. Also Sandeep kaur is having 8 publications in various National and International Conferences.

**Er. Anantdeep**, presently working as Lecturer in the Department of CSE/IT, BBSB Engineering College, Fatehgarh Sahib. She is M-Tech in ICT from Punjabi university Patiala. Her major research interests include mobile zigbee networks. Also Anantdeep is having 3 publications in various National and International Conferences.

**Er. Deepak Aggarwal**, presently working as Lecturer in the Department CSE/IT of BBSB Engineering College, Fatehgarh Sahib,Punjab (INDIA). He is having a total teaching experience of about 7 years & he has done Master of Technology from Punjab Technical University. His major research interests include DIP and performance evaluation of networks. Also Deepak Aggarwal is having to his credit about 8 publications in various National and International Conferences and Journals.